\documentclass[preprint, showpacs, amsmath,amssymb]{revtex4}

\usepackage{graphicx}
\usepackage{dcolumn}
\usepackage{bm}

\begin{document}
\preprint{pre-print number}
\title{General multistate models for agents with internal bias}
\author{Juan G. Diaz Ochoa}  
\affiliation{\mbox{Institute of Physics, Fachbereich 1, University of Bremen,
Otto Hahn Alle, D-28334 \ Bremen, Germany}}
\email{diazochoa@itp.uni-bremen.de} 
\author{Elena Ramirez Barrios}
\affiliation{\mbox{Institute of
Economics, Fachbereich 7, University of Bremen, Hochschulring 4, D-28359 \ Bremen,
Germany}}

\begin{abstract}
We present a model of interpersonal comparisons appearing as a generalization of a multi-state model for elements with internal bias. Within this model agents suffering under dissatisfaction compare them with their neighbors. The internal bias are the pre-formed preferences, and the states are represented by the visual contacts that an agent makes to the baskets of her/his neighbors. The topology of the comparisons is a random network. We compare the behavior between altruistic and non-altruistic agents and we find out that altruistic behavior alters the robustness of the network.
\end{abstract}

\maketitle



\section{Introduction}

Complex networks connected by web like structures have been intensively studied, assuming that their elements are homogeneous. Such networks could be generated by means of technical elements and be used for communication \cite{BarabasiI} (airports, wireless links, internet, etc.) or could be defined as communication between individuals (social networks) \cite{Milgram}.

We model agents that, after a distribution of goods, subtend a network of interpersonal comparison. The fair distribution of goods is an old problem in mathematics that concerns not only distribution of perfectly divisible goods (for example a cake), but also the distribution of for example $CO_{2}$ emissions \cite{Steinhaus,Brams}. Along this work we do not search for an optimal distribution, but analyse the 'feelings' arising after this distribution is done. After a distribution takes place, agents are allowed to make interpersonal comparisons, depending on how this distribution was done and how were the individual satisfaction levels among the agents \cite{Elena1}. The mechanism is as follows: each agent is endowed with a pre-established list of preferences that produce a bias in the distribution. Each agent tries to optimize the number of elements he/she gets according to this preference list. This bias appears not only as a pre-distribution but also a post distribution mechanism: the agents are allowed to subtend a network of interpersonal comparisons according to the preferences that were not fulfilled. Such a network is basically an information-exchange network. This model can be generalized for information networks that emerge in systems which are not symmetric and far from equilibria, because of the individual bias.

The existence of individual bias (preferences) could also change the topology of the network. This characteristic is relevant for weighed unidirectional networks \cite{Cajueiro}. However, the general alternative proposed in this work may motivate a new class of models for random networks. The nodes of our network are not only defined by the agents, but also by the material and cognitive possessions each agent has. Each state is not aprioristically defined, but is computed after the distribution process, depending on the individual bias. At difference with conventional models (for a general review see \cite{BarabasiI}), our approach can concern the definition of elements (nodes of the comparison network), not only with different but also with changing identities. 

An important task concerning random networks is to describe how its formation and connectivity are. For instance, the analysis of the distribution of connections per node has been analyzed for Internet \cite{Barabasi}. In a previous work the connectivity was only analyzed as a function of the kind of distribution \cite{Elena1}. But agents, making interpersonal comparisons, are also exchanging information in form of visual contacts. This information can randomly be controlled, for example, avoiding that an agent look into the basket of other agent in the neighbor. We call this mechanism censure. This control is equivalent to a random breakdown, where elements in the network are randomly removed \cite{BarabasiI, Cohen}. If $p$ is a fraction of nodes under censure, then, after some threshold value $p_{c}$, the network disintegrates it into smaller disconnected parts. Below the critical threshold value the network, connecting each of the nodes, persists. The random breakdown of a network trespassing the threshold of an infinite dimensional percolation problem. 

Non homogeneous and constant nodes have been already studied \cite{Cheng}. As a difference our nodes also are able to change their identities, making the inhomogeneities dependent on the time. The agents can behave in two different ways. Either they simply subtend a network of interpersonal comparisons after the distribution process, or they can be altruistic. Altruistic agents are giving some of their owned goods to agents with a lower satisfaction level, attempting to improve the welfare of the system. Each node consists of a cognitive part (the agent able to make decisions and with a mental bias) and a material part (the basket with goods). Thus, a donation of a good implies an identity change. By using this example we try to answer the question, if this change of identity may induce a change in the percolation transition (breakdown of the connectivity of the nodes) of the network. 

\section{Model and formalism}

\emph{Hamiltonian:} The internal restrictions imposed to individual agents are the basis of a non-equilibrium state. For instance, in the case of a distribution of goods, individual bias does not allow an equitable distribution. However, the system tries to compensate this non-equilibrium state by means of interpersonal comparisons that could later imply interpersonal exchange of goods. This situation has been previously implemented for individual agents randomly matched according to similarity of preferences promoting exchange \cite{Axelrod}. These comparisons are essentially an exchange of information, because each agent gets 'visual' information of the goods other agents own. Each agent has a bias to the kind of goods but not to that neighboring agent. Therefore, this information exchange takes place in a random way, conforming a random network between agents. In the following part the definition of the Hamiltonian for the exchange and its states is done.

For each interpersonal comparison we define a Hamiltonian in a similar way as for a multi-state system, where the energy relation of the system is given by its connectivity \cite{Johannes}. The function $E_{i}(j)$ depends on the order of the restriction imposed to the system. Here the restriction is the preference ranking in a system of envious agents. If it is not possible to establish any comparison between two agents with the first element of the list, then the second element in the list is used as restriction to the comparison. We define here this restriction by $C^{k}_{i}$ in the following form
\begin{equation}
C^{k}_{i}=\begin{cases}1, & \text{if agent $i$ is still in doubt to be satisfied or not before $k$ comparisons take place}\\ 0 & \text{otherwise}\end{cases}.
\end{equation}
An interpersonal comparison is allowed only if an agent express some dissatisfacion. This expression is more precisely defined in the following way  
\begin{equation}
C^{k}_{i}=\prod_{l=1}^{k-1} \delta(b_i(P_i(l)),b_j(P_i(l)));
\label{Restr_pref}
\end{equation}
in eq. (\ref{Restr_pref}) $b_{i}$ are the baskets, where the agents store their goods after the distribution. $P_{i}(l)$ is the preference list, that depends on the class of goods $l$. Using this notion, the function $E_{i}(j)$ has the following form
\begin{equation}
E_{i}(j)=\sum_{k=1}^{K}\Theta (b_{j}(k)-b_{i}(k))C^{k}_{i},  
\label{State}
\end{equation}
where $\Theta (x)$ is the step function, $b_{i}(k)$ represents the number of elements of type $k$ in the basket of agent $i$, where $k$ is the number of elements into the basket $b_{i}$ of each agent $i$ and $K$ the total good types. From equation (\ref{State}) we are able to derive the following cases
\begin{equation}
E_{i}(j)=\begin{cases}1 & \text{if agent $i$ is not satisfied (feels envy)}\\ 0 & \text{otherwise} \end{cases}.
\end{equation}
The equation (\ref{State}) is the definition of the individual states of the system. The Hamiltonian is in general defined via the connectivity between the states of the different agents, and is given as the sum over $i$ and $j$ of Eq. (\ref{State}) 
\begin{equation}
E=\sum_{i=1}^N \sum_{j=1}^N \eta(i,j) \times E_{i}(j).
\end{equation}
This model belongs to a general class of models, which under some constrains appear like a conventional multi-state system \cite{Binder}.

\emph{Dynamics:} One agent $i$ is linked to the agent $j$ if the state $E_{j} > 0$. The transition probability is given by 
\begin{equation}
W_(t;t+1) = \Theta(E_i(t)-E_{c}),
\end{equation}
with $E_{i}=\sum_{j=1}^N \eta(i,j) \times E_i(j)$. In this expresion, $\Theta(E_i(t)-E_{c})$ is again the step function. In our implemented model, the treshold value is given by $E_{c}=0$. The updating for the number of envious agents is given by a Langevin equation of motion \cite{Jensen}, given by

\begin{equation}
N_{i}(t+1) = N_{i}(t) - N_{i}(t)W_{N_{i}}(t;t+1)  +\sum_{i=1}^N N_{i} W_{N_{i}}(t;t+1).
\label{Langev}
\end{equation}
In a previous study, the identity of the elements has been defined in a static form \cite{Elena1}. However, this identity can change and be tracked. This assumption is more realistic to model multi-agent systems, than the consideration of fixed properties. We assume a pre-distribution inside the states $b_{i}$ previous to the conformation of the connections between states. In our example of dissatisfied agents, this assumption is equivalent to altruistic agents that donate the elements of their baskets to other agents.

Indeed, an additional equation of motion for the redistribution of the elements among the states $b_{i}$ must be considered. A donation from agent $j$ to agent $i$ is given by 
\begin{equation}
b_{i}(t+1)=b_{i}(t) + \sum_{j=1}^{N_{d}} \delta (b_{i}-b_{j})\delta (\omega
-\omega _{0}),
\label{Altr_1}
\end{equation}
whereas for the agent $j$ the following process (substraction inside the state $b_{j}$) takes place 
\begin{equation}
b_{j}(t+1)=b_{j}(t) - \sum_{i=1}^{N_{d}} \delta (b_{i}-b_{j})\delta (\omega
-\omega _{0});
\label{Altr_2}
\end{equation}
in equations (\ref{Altr_1}) and (\ref{Altr_2}) $N_{d}$ represents the number of donators into the system and $\omega_{0}$ is the allowed donation frequency. Both equations of motion (\ref{Altr_1}) and (\ref{Altr_2}) are coupled to the equation of motion for the number of connected states in Eq. (\ref{Langev}).

The whole system is subjected to an initial non-symmetrical distribution $G(K)$ of $K$ elements among the agents. As was defined in a similar model \cite{Elena1}, where the distribution function were defined as a gamma function, the symmetry of the distribution of goods among the individuals depends on a parameter $A$, i.e if $A \rightarrow 0$, then the distribution is not fair. This asymmetry sets the system out from equilibrium and motivates the interpersonal comparison. In this context a question remains: is the information exchange, the initial distribution, defined by $A$, or both parameters relevant for the behavior of the system? Which of both parameters can modulate the interpersonal comparisons into the system?

\section{Results}

The exchange of information between agents (or agent connectivity) depends on the parameter $\kappa$ which is related to $\eta_{i,j}$ as $\kappa= \sum_{i<j}\eta _{i,j}$, which can be understood by means of the following aphorism: the social temperature of a system depends on the censure in the system. With less censure the agents are allowed to compare themselves, increasing the fluctuations of the system. In the present results, $\kappa$ represents the censure (temperature) parameter, which is also equivalent to the breakdown parameter of the network. 

\begin{figure}[th]
\begin{center}
\includegraphics[clip,width=0.4\textwidth]{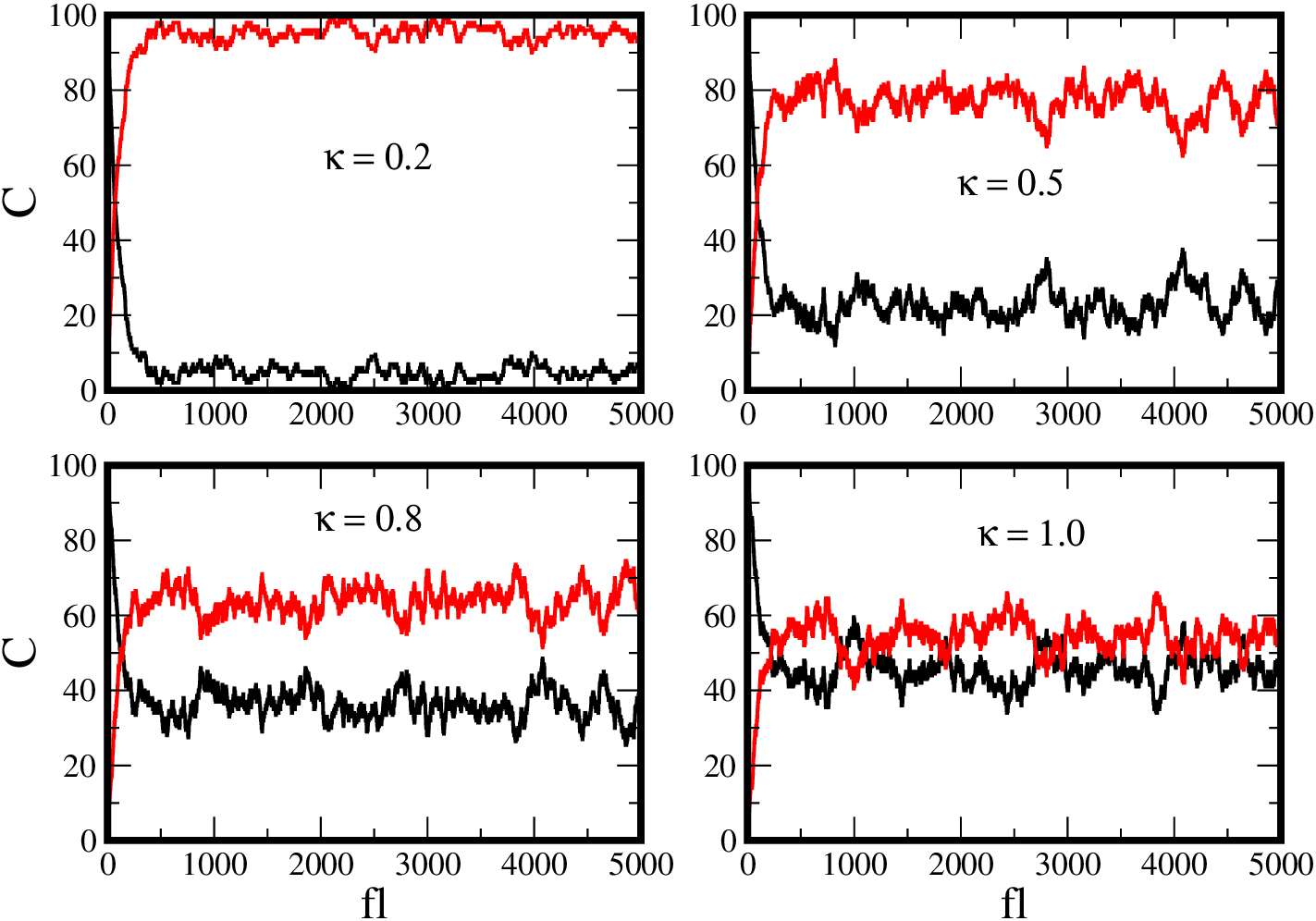}
\caption{Time series of the number of agents with and without interpersonal comparisons for four parameters $\kappa$ ($\kappa = 0.1$ upper left, $\kappa = 0.5$ upper right, $\kappa = 0.8$ under left, $\kappa = 1$ under right) and fixed initial distribution $A = 80$. By increasing $\kappa$ the number of linked agents approaches to the number of non-linked agents.}
\label{fig1}
\end{center}
\end{figure}

The control on the connectivity is equivalent to the random breakdown of networks, i.e. the number of agents that are not able to exchange information is equivalent to remove the agent from the network. The principal observables here are the fraction of nodes that resist to interpersonal comparisons as a function of $\kappa$. Using this notion, the number of linked states $C$ as a function of the frequency of information exchange $fl$ was estimated. For $fl = 0$ the system starts from a fully non-interconnected state. Whereas the system evolves, the connectivity between agents increases. For $fl \rightarrow \infty$ the system reaches an equilibrium. The time series for some states as a function of $\kappa$ is shown in Fig. \ref{fig1}.

\begin{figure}[th]
\includegraphics[clip,
width=0.4\textwidth]{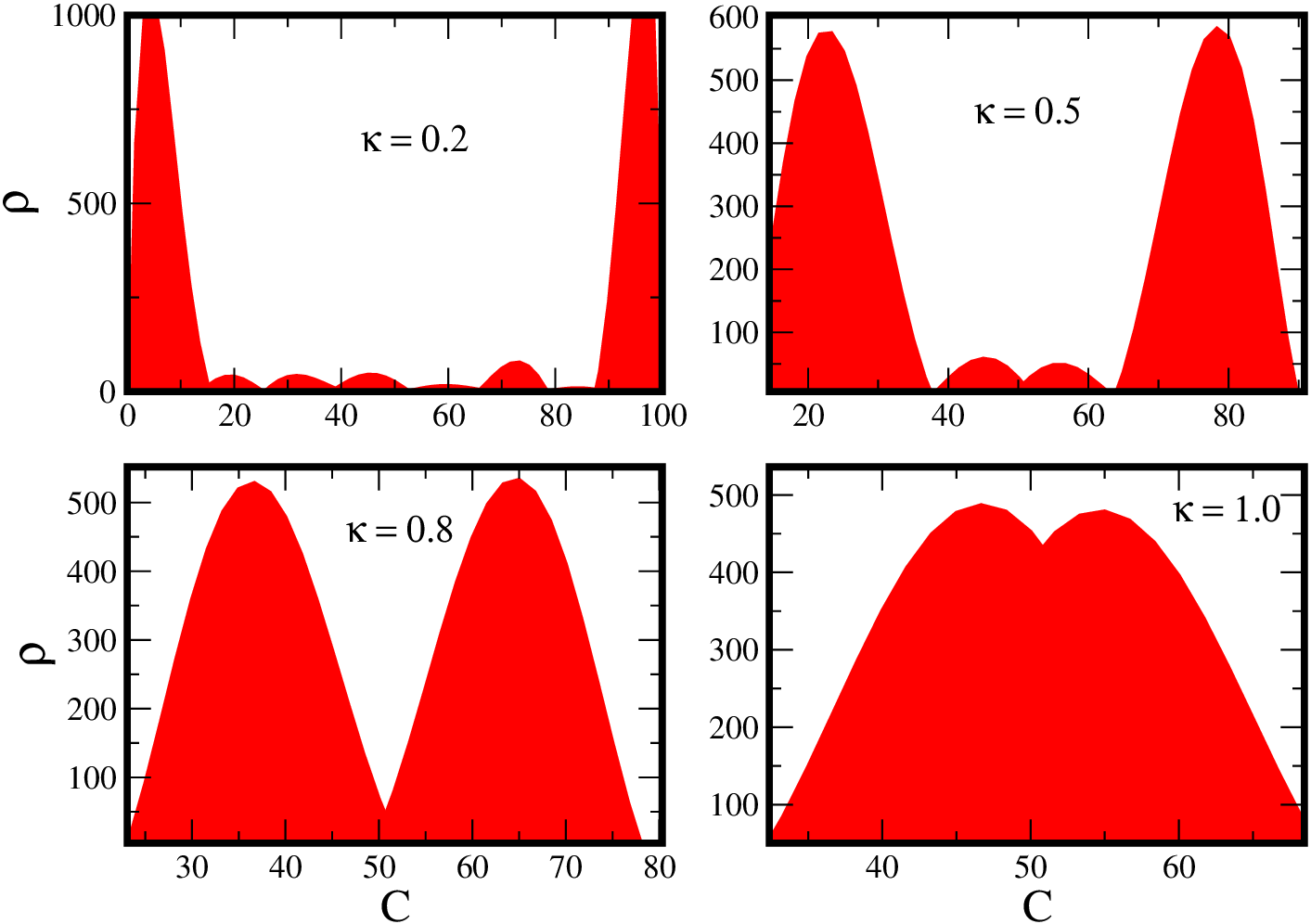}
\caption{Non normalized density of states $\rho(C)$ is presented for different parameters $\kappa$ ($\kappa = 0.1$ upper left, $\kappa = 0.5$ upper right, $\kappa = 0.8$ under left, $\kappa = 1$ under right) and fixed initial distribution $A = 80$. Due that there are agents that reach a high satisfaction level in few iteration steps, there is no total mixture of satisfied and unsatisfied agents at $\kappa \sim \kappa_{c}$.}
\label{fig2}
\end{figure}

For very small values of $\kappa$, there is a clear separation of linked and non-linked agents. Conform the connectivity increases, the phase separation decreases. Eventually at some $\kappa_{c}$ there is a mixture of linked and non-linked agents. For values above this point the system saturates, i.e. reaches a maximum in the number of links between agents. The number of linked states $C$ is shown in Fig. \ref{fig2}. Here it is important to observe that there is no complete mixture between linked and non-linked agents for $\kappa > \kappa_{c}$. Given that in each iteration there are agents that are completely satisfied, they are of course indifferent to the modification of the parameter $\kappa$, yielding a shift in the distribution of linked agents.

\begin{figure}[th]
\begin{center}
\includegraphics[clip, width=0.5\textwidth]{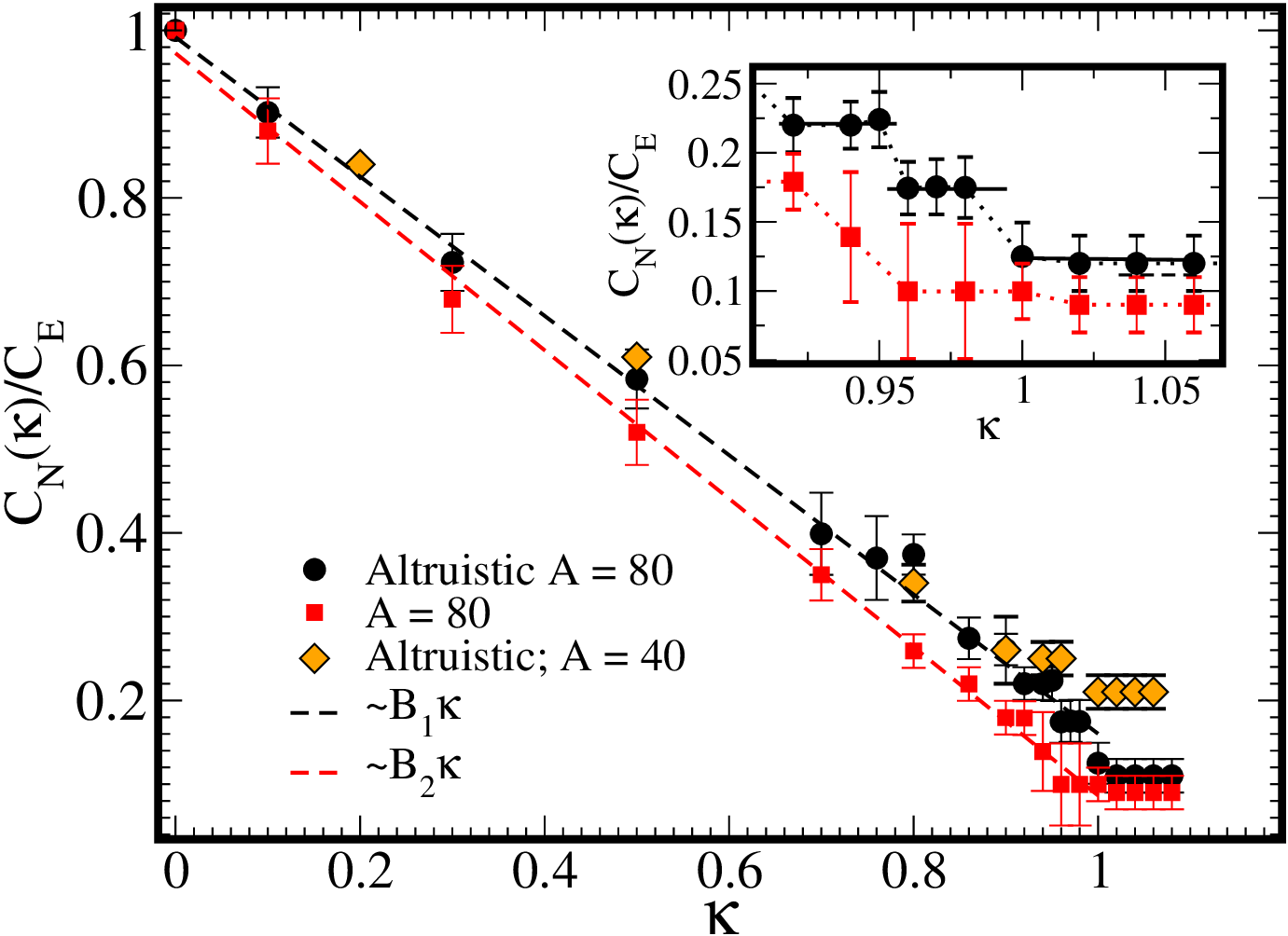}
\end{center}
\caption{Phase diagram of the fraction of agents resistant interpersonal comparisons as a function of the random breakdown of the network, given by the parameter $\kappa$. The behavior for both altruistic and non-altruistic agents is compared; both cases were fitted with a linear regression, with two slopes $B_{1}$ and $B_{2}$. The behavior of altruistic agents for $A=40$ is also shown. In the inset the behavior of the curve close to the critical parameter $\kappa _{c}$ is shown.}
\label{figA}
\end{figure}

The role of $\kappa$ is decisive for the transition from fully interconnected agents to isolated ones. The phase behavior of the density of states as a function of $\kappa$ has been investigated and plotted in Fig. \ref{figA}. Indeed, two main cases are compared: when an initial distribution $G(K)$ and interpersonal comparisons arises, and when an additional altruistic distribution of goods between agents takes place. This comparison is made for a fixed distribution $A$. In this plot the fraction of the number of the non connected nodes in relation to the connected nodes $C_{N}=(C-C_{E})/C_{E}$ as a function of $\kappa$ (normalized) is shown, where $N$ is the maximal number of connected nodes.

In both cases, the relative density of states decays in a lineal form, i.e. $C \sim \kappa$. The slope of the curves depends on whether there is altruistic exchange or not. This lineal dependence is valid for $\kappa << \kappa_{c}$. However, conform $\kappa$ approaches to its critical value, a non-linear behavior for values of $C_{N}/C_{E}$ is found. For non altruistic agents there is a small step for $0.92 < \kappa < 0.96$. For $\kappa > 0.96$ the system reaches a saturation, and the relative density reaches a plateau. This behavior could be related with finite size effects.

The presence of altruistic agents makes the behavior of the system more interesting. The phase diagram has at least two plateaus for $\kappa < \kappa_{c}$, one at $\kappa=0.92$ and $\kappa=0.96$. The last plateau is reached when the system is saturated, i.e. for $\kappa \le 1$. One can assume that this effect is due to statistical errors in the estimation of the results. However, the error estimated is above the curve for non-exchange (see the inset of Fig. \ref{fig2}). Despite the network presents these breakdowns, the network still preservesits robustness while $\kappa$ increases. The transition in this case takes place for $\kappa$ larger than the transition without exchange. 

If altruism takes place, the interconnectivity cannot increase and an internal resistance grows until a new saturation is reached. Indeed, this behavior can be described as a kind of fractional formation of interconnected agents, where exchange between agents on $k$ changes the internal energy of the system.

One can suppose that the initial distribution $G(k)$ may change the phase behavior of the interconnectivity between agents. The simulations performed for $A=40$ show that the behavior is similar to $A=80$ for $\kappa < \kappa_{c}$. This result confirms and generalizes the results obtained in a previous work \cite{Elena1}. For $\kappa \sim \kappa_{c}$ the fraction of agents resistant to interpersonal comparisons is higher. This result is surprising but not strange, because the more asymmetric the initial distribution the faster some agents can fill their baskets, increasing the number of agents that are insensitive to the parameter $\kappa$. In both cases $\kappa_{c}$ is a fixed point. Hence, the symmetry of the initial distribution modifies the phase behavior of the system near the critical point, but not the whole behavior, which essentially depends on the information exchange between agents and on the change of the states $b_{i}(k)$ by internal redistribution of $k$, i.e. on information exchange and altruism. 

\section{discussion}

In general, the change of identity of the elements of a random network may influence its robustness. We were not able to introduce a dramatic change in the robustness of the network; however, it is possible that extreme fluctuations in the identity of the elements may introduce changes, increasing the resistance to make new connections in the network. In extreme cases such fluctuations can break down the connectivity of the network. Another important aspect is the definition of the network. In specific problems, the identity of the elements may depend on initial definitions that cannot be made ad-hoc, as has been shown in the present example of interpersonal comparisons.

Intuitively one can assume that altruism, or redistribution on the number of objects $k$ among the agents, can avoid the formation of interconnected states. However, the present results show that above a critical censure parameter $\kappa_{c}$ the system is fully interconnected. The effect of altruism, as defined for the present work, is remarkable for states close to $\kappa_{c}$: it does not avoid the transition from connected to disconnected states, but offers a resistance in the formation of such connections

The modulation of the information exchange, i.e., the intensity of the censure, and not the distribution, is essential in the control of interpersonal comparisons. However, the symmetry of the elements and their equilibrium, which depends, in the present example, on the distribution of goods among the agents, have influence in the robustness of the network. If the system is close to equilibrium, i.e. there is an equitable distribution (when the parameter $A$ is large), the agents have more chance to make donation of goods because there are more available goods to give, modifying the frequency of interpersonal comparisons. Otherwise, the asymmetry of the initial distribution cannot be prevented by donations and a network of comparisons emerges.

In the whole model the individual biases play a relevant role. They are relevant for the definition of part of the nodes, of the interpersonal comparisons and the promoting of an altruistic behavior. Hence the formation of these biases is relevant in the formulation of such kind of models. The simple variant adopted in this work is to define them in a random way. However, information exchange should be also considered in more realistic models, not only as simple interpersonal comparison, but also in formation and modification of such biases. 

{\bf Aknowledgments} We want to thank Johannes J. Schneider (University of Mainz) for the initial formulation of the simulation code and his help in the suggestion of the initial theoretical background implemented in this work.


\end{document}